\documentclass[12pt,a4paper]{article}
\usepackage[text={16.5cm,23cm},centering]{geometry}
\usepackage{amsmath,amssymb}
\usepackage[utf8]{inputenc}
\usepackage{tgtermes,tgheros,tgcursor}
\usepackage[varg]{newtxmath}
\usepackage{hyperref}
\usepackage[boxsize=0.8ex]{ytableau}
\usepackage{xspace}

\newcommand{\del}{\partial}
\newcommand{\U}{\mathrm{U}}
\newcommand{\SU}{\mathrm{SU}}
\newcommand{\su}{\mathrm{su}}
\newcommand{\Zb}{\mathbb{Z}}
\newcommand{\yasym}{\;\raise1ex\hbox{$\ydiagram{1,1,1}\,$}\xspace}
\newcommand{\psib}{\bar{\psi}}
\newcommand{\sigmab}{\bar{\sigma}}
\newcommand{\Acal}{\mathcal{A}}
\newcommand{\Fcal}{\mathcal{F}}
\newcommand{\id}{\vvmathbb{1}}

\DeclareMathOperator{\tr}{\mathrm{tr}}
\DeclareMathOperator{\Tr}{\mathrm{Tr}}

\begin{document}
\begin{center}
  \begin{flushright}
    OU-HET 988
  \end{flushright}
  \vspace{8ex}
  {\Large \bfseries \boldmath 't~Hooft anomaly matching condition and chiral symmetry breaking without bilinear condensate}\\
  \vspace{4ex}
  {\Large Satoshi Yamaguchi}\\
  \vspace{2ex}
  {\itshape Department of Physics, Graduate School of Science, 
  \\
  Osaka University, Toyonaka, Osaka 560-0043, Japan}\\
  \vspace{1ex}
  \texttt{yamaguch@het.phys.sci.osaka-u.ac.jp}\\
  \begin{abstract}
  We explore 4-dimensional SU$(N)$ gauge theory with a Weyl fermion in an irreducible self-conjugate representation. 
  This theory, in general, has a discrete chiral symmetry.  
  We use 't~Hooft anomaly matching condition of the center symmetry and the chiral symmetry, and find constraints on the spontaneous chiral symmetry breaking in the confining phase.  
  The domain-walls connecting different vacua are discussed from the point of view of the 't Hooft anomaly.
  We consider the SU$(6)$ gauge theory with a Weyl fermion in the rank 3 anti-symmetric representation as an example.  
  It is argued that this theory is likely to be in the confining phase.  
  The chiral symmetry $\mathbb{Z}_6$ should be spontaneously broken to $\mathbb{Z}_2$ under the assumption of the confinement, although there cannot be any fermion bilinear condensate in this theory.
  \end{abstract}
\end{center}

\vspace{4ex}
\section{Introduction}

't~Hooft anomaly matching condition \cite{tHooft:1979rat} is a powerful tool to investigate the phase structure of strongly coupled quantum field theories.  
Since the 't~Hooft anomaly is invariant under the renormalization group flow, it often strongly constrains the vacuum structure and the low energy effective theory.  
't~Hooft anomaly matching condition of discrete symmetry has also been shown to be useful \cite{Csaki:1997aw}.

Recently it has been shown in \cite{Gaiotto:2014kfa,Gaiotto:2017yup} that 't~Hooft anomaly matching including higher form symmetries is also powerful.  
In particular the center symmetry of the gauge theory, a typical example of a 1-form symmetry, and its twist are related to confinement; 
the confining phase is characterized by the absence of spontaneous breaking of the center symmetry. 
Therefore by considering the mixed 't~Hooft anomaly of the center symmetry,  one may find constraints of the phase structure of the theory in the confining phase.
There has been interesting progress along this line \cite{Tanizaki:2017bam,Komargodski:2017smk,Shimizu:2017asf,Gaiotto:2017tne,Kitano:2017jng,Yamazaki:2017dra,Cordova:2017vab,Tanizaki:2017mtm,Cherman:2017dwt,Cordova:2017kue,Hung:2018rhg,Draper:2018mpj,Anber:2018tcj,Cordova:2018acb,Anber:2018jdf,Tanizaki:2018wtg,Bi:2018xvr}.

In this paper we investigate a 4-dimensional $\SU(N)$ gauge theory with a Weyl fermion in an irreducible self-conjugate representation $R$. 
The $\U (1)$ phase rotation of the fermion in this theory is broken to $\Zb_{\ell},\  (\ell: \text{ Dynkin index of } R)$ due to the anomaly \cite{Adler:1969gk,Bell:1969ts}.  
We call this $\Zb_{\ell}$ symmetry the ``chiral symmetry.''  
In addition, this theory has the $\Zb_q$ center symmetry, where $q:=\gcd(N,c)$  and $c$ is the $N$-ality of $R$.  
We check the mixed 't~Hooft anomaly between the chiral symmetry and the center symmetry, and obtain constraints on the spontaneous breaking of the chiral symmetry under the assumption of confinement. 
In this analysis we assume that a gauge theory does not reproduce the 't Hooft anomaly, if the theory is in the confining phase and the global symmetries are not spontaneously broken.  

For example, in the $\SU(6)$ gauge theory with a Weyl fermion in the rank 3 anti-symmetric representation denoted by the young diagram \yasym, the chiral symmetry $\Zb_6$ is broken to $\Zb_2$ if it is in the confining phase.  
By comparing this theory and $\mathcal{N}=1$  $\SU (6)$ pure super Yang-Mills theory, we find that this theory is quite likely to be in the confining phase, and thus the chiral symmetry is quite likely to be broken spontaneously.

Interestingly the above theory cannot have fermion bilinear condensate, since a Lorentz invariant and gauge invariant fermion bilinear vanishes identically in this theory.  
Therefore this theory is an example in which the chiral symmetry is spontaneously broken without fermion bilinear condensate.  Such possibility or impossibility has been discussed in QCD \cite{Kogan:1998zc,Kanazawa:2015kca,Tanizaki:2018wtg} as well as the gauge theory with adjoint fermions \cite{Anber:2018tcj,Cordova:2018acb}.  It has been argued that the chiral symmetry breaking without fermion bilinear in the $\SU (2)$ gauge theory with a Weyl fermion in spin $3/2$ representation \cite{Poppitz:2009kz}, in which the fermion bilinear vanishes identically.

In order to break the chiral symmetry, a four-fermi operator is likely to condense.  
There are two four-fermi operators which are gauge invariant, Lorentz invariant, and charged under the chiral symmetry.  
It will be an interesting future problem to find which four-fermi operator condenses.
Other candidates which causes this spontaneous symmetry breaking are fermion bilinears including derivatives.  
We find that a fermion bilinear with two derivatives cannot condense.  
Thus only a fermion bilinear with four or more derivatives may condense.

We also discuss dynamical domain-walls connecting distinct vacua, which appears when the discrete chiral symmetry is spontaneously broken.  
We find that the domain-wall supports a conformal field theory or a topological field theory which reproduce the 't Hooft anomaly for the center symmetry by a similar argument to \cite{Gaiotto:2014kfa}.

The construction of this paper is as follows. 
In section \ref{sec:anomaly}, we give the detail of the setup and consider the 't~Hooft anomaly matching condition.  
Domain-walls are also discussed.
In section \ref{sec:example}, we discuss the SU(6) gauge theory with a Weyl fermion in \yasym.  
In particular we argue that this theory is in the confining phase and thus the chiral symmetry is broken spontaneously. Section \ref{sec:summary} is devoted to summary and discussion.

\section{'t~Hooft anomaly matching condition}
\label{sec:anomaly}

\subsection{Setup}
We consider an $\SU(N)$ gauge theory with a Weyl fermion in an irreducible self-conjugate representation $R$.  Let us first summarize some group theoretical conventions.  It should be useful to consult \cite{Slansky:1981yr,Yamatsu:2015npn}.  Let $T_a,\ a=1,\dots, N^2-1$ be the generators of the Lie algebra $\su(N)$ normalized as
\begin{align}
  \tr[T_a T_b]=\frac12 \delta_{ab},
\end{align}
where $\tr$ is the trace in the fundamental representation.  The Dynkin index $\ell$ for an irreducible representation $R$ is defined by\footnote{$\ell$ in this paper is the same one as $l(\Lambda)$ in \cite{Slansky:1981yr} and $2T(R)$ in \cite{Yamatsu:2015npn}.}
\begin{align}
  \Tr_{R}[T_a T_b]=\ell \tr[T_aT_b],
\end{align}
where $\Tr_{R}$ denotes the trace in the representation $R$.  The Dynkin index $\ell$ is known to be a non-negative integer.  Another important quantity is the $N$-ality, or the conjugacy class $c$, defined as the number of boxes modulo $N$ when $R$ is expressed by a Young diagram for $\su(N)$.

Next let us turn to the fields and the action of our theory.  The gauge field is denoted by $A=A_{\mu}dx^{\mu}=A_{\mu}^aT_a dx^{\mu}$.  
The field strength $F$ is defined by $F=dA-i A\wedge A=\frac12 F_{\mu\nu}dx^{\mu}\wedge dx^{\nu}$.  
The fermion is denoted by $\psi$ and $\psib$ whose components are $\psi_{\alpha}^{I},\psib_{\dot{\alpha}I}$, where $\alpha,\dot{\alpha}=1,2$ are two component spinor indices, and $I=1,\dots,\dim R$ is the ``color'' index. 
The action in the Euclidean signature is written as
\begin{align}
  S=\int \frac{1}{g^2}\tr[F\wedge *F] + \int d^4x i\psib \sigmab^{\mu}D_\mu\psi,\qquad 
  D_{\mu}\psi:=\del_{\mu}\psi-iA_{\mu}^{a}T_{a}^{R}\psi,
  \label{action}
\end{align}
where $g$ is the gauge coupling constant, $T_a^{R}$ is the representation matrix of $T_a$ in $R$, and $\sigmab^{\mu},\mu=1,2,3,4$ are $2\times 2$ matrices defined in \cite{Wess:1992cp} besides $\sigmab^{4}:=i\sigmab^{0}$.

We limit ourself to the case where $R$ is self-conjugate (real or pseudo-real) so that our theory is free from perturbative gauge anomaly since the anomaly coefficient $\Tr_{R}[T_a\{T_b,T_c\}]=0$ if $R$ is self-conjugate.
We also limit ourself to the case with $N>2$ so that our theory is free from the global gauge anomaly of \cite{Witten:1982fp}.  

\subsection{Chiral symmetry}
The action is invariant under the $\U (1)$ rotation by which $\psi$ and $\psib$ are transformed as
\begin{align}
  \psi\to e^{i\alpha}\psi,\quad
  \psib\to e^{-i\alpha}\psib,\qquad
  (\alpha:\text{ constant parameter}).
  \label{axialUone}
\end{align}

However this symmetry is broken quantum mechanically by the Adler-Bell-Jackiw anomaly \cite{Adler:1969gk,Bell:1969ts}.  
This anomaly is the change of the path-integral measure due to Fujikawa \cite{Fujikawa:1979ay}.  
By the transformation \eqref{axialUone}, the path-integral measure changes as
\begin{align}
  \int D\psi D\psib \to \int D\psi D\psib \, e^{i\alpha \ell \nu},\quad
  \nu:=\frac{1}{8\pi^2}\int \tr(F\wedge F),
  \label{ABJanomaly}
\end{align}
where $\ell$ is the Dynkin index of the representation $R$.

In eq.~\eqref{ABJanomaly}, since the instanton number $\nu$ is an integer, 
the path-integral measure is invariant as well if $\alpha$ is written as
\begin{align}
  \alpha=\frac{2\pi n}{\ell},\quad n\in \Zb.
\end{align}
Therefore the $\Zb_{\ell}$ symmetry given by the following equation exists even in the quantum theory.
\begin{align}
  \psi \to e^{2\pi i n/\ell} \psi,\quad
  \psib \to e^{-2\pi i n/\ell} \psib,\qquad
  n\in\Zb.
  \label{chiralsymmetry}
\end{align}
We call this symmetry the ``chiral symmetry.''

In this paper we discuss the spontaneous breaking of this $\Zb_{\ell}$ chiral symmetry.

\subsection{Center symmetry}
Let us discuss the center symmetry of our theory.  The center symmetry group of the pure $\SU (N)$ Yang-Mills theory is $\Zb_{N}$.

It is convenient to use the lattice gauge theory picture.  In the lattice gauge theory, the gauge field is expressed by the link valuable $U_{r}\in \SU (N)$ associated to each link $r$.  The pure Yang-Mills action $S_{G}(U)$ is written in terms of ``plaquettes'' or any other ``small'' Wilson loops.

Let us first see how the center symmetry is realized in the lattice gauge theory.  
First we choose an oriented co-dimension one surface $\Sigma$ in the space-time which does not cross the lattice sites.  
Then for an element $z\in \Zb_N$, the transformation of the link valuables is given by
\begin{align}
  U_{r}\to
  \begin{cases}
    z U_{r} & (r \text{ intersects } \Sigma \text{ positively}), \\
    z^{-1} U_{r} & (r \text{ intersects } \Sigma \text{ negatively}), \\
    U_{r} & (r \text{ does not intersect } \Sigma). \\
  \end{cases}
  \label{centertransformation}
\end{align}
Since total intersection number of a plaquette and $\Sigma$ is $0$, it is invariant under this transformation.  Any small Wilson loop is also invariant for the same reason.  Therefore the action $S_G(U)$ is invariant under this transformation \eqref{centertransformation}.

Next let us consider the fermion part.  We assume the fermion part of the action is written as $S_F(R(U),\psi,\psib)$ where $R(U)$ is the representation matrix of $U$ in $R$.
If $R(zU)=R(U)$ for $z\in \Zb_N$, then $S_F$ is invariant under the transformation \eqref{centertransformation}.  For $z=e^{2\pi i m/N}, \ (m\in \Zb$), $R(zU)=R(U)$ implies
\begin{align}
  e^{2\pi i m c/N}=1,\qquad (c:N\text{-ality}),
\end{align}
since $R(z)=z^c$.
This further implies
\begin{align}
  mc=Nk,\qquad \exists k \in \Zb.
  \label{mcNk}
\end{align}
Let $q=\gcd(c,N)$ and $N=qN_0,\ c=qc_0$. Then $N_0$ and $c_0$ are relatively prime, and eq.~\eqref{mcNk} is rewritten as
\begin{align}
  mc_0=kN_0.
\end{align}  
This is solved as
\begin{align}
  m=N_0 n,\quad k=c_0 n,\qquad n\in \Zb.
\end{align}
As a result, $S_F$ is invariant under the transformation \eqref{centertransformation} if $z=e^{2\pi i N_0 n /N}=e^{2\pi i n /q} \in \Zb_{q}\subset \Zb_{N}$.

In summary, our theory has $\Zb_{q}$ center symmetry.  Here $q=\gcd(c,N)$ and $c$ is the $N$-ality of $R$.

Actually since $R$ is a self-conjugate representation, the $N$-ality $c$ is $0$ or $N/2$.  $c=N/2$ is possible only when $N$ is even.  
$q:=\gcd(c,N)$ is given by $q=N$ when $c=0$ while $q=N/2$ when $c=N/2$. 
Since we concentrate on the case $N>2$, $q$ is always greater than $1$ and thus our theory has non-trivial center symmetry.

We can relate this center symmetry to the confinement as usual.  Actually the Wilson loop in the fundamental representation exhibits area law if this center symmetry is not spontaneously broken. 

\subsection{'t Hooft anomaly}

Let us discuss mixed 't~Hooft anomaly between the center symmetry and the chiral symmetry, and find some constraints on the spontaneous chiral symmetry breaking.  
Here we look at non-invariance by the chiral symmetry transformation when the background center symmetry gauge field is introduced as in \cite{Gaiotto:2017yup,Gaiotto:2017tne,Shimizu:2017asf}.

We formulate the gauge field for the 1-form $\Zb_q$ symmetry following \cite{Gaiotto:2017yup,Kapustin:2014gua,Gaiotto:2014kfa}.  
First we introduce 2-form $\U (1)$ gauge field $B$ and 1-form $\U (1)$ gauge field $C$.  They are normalized such that $\frac{1}{2\pi}\int_{2\text{-cycle}}dC$ and $\frac{1}{2\pi}\int_{3\text{-cycle}}dB$ take arbitrary integer values.  Then we relate $B$ and $C$ by the constraint
\begin{align}
  qB=dC.\label{constraint}
\end{align}
The gauge transformation of this $(B,C)$ is given by
\begin{align}
  B \to B+d\lambda,\qquad
  C \to C+q\lambda+df,
  \label{2formgauge}
\end{align}
where a gauge transformation parameter $\lambda$ is a $\U (1)$ gauge field, and another gauge transformation parameter $f$ is a periodic scalar field with identification $f\sim f+2\pi$.  Then $B$ is almost pure gauge because of the constraint \eqref{constraint}.  However the Wilson surface $s=\exp\left(i\int_{\text{2-cycle}}B\right)$ can take non-trivial value.  Actually taking the constraint \eqref{constraint} into account $s^q$ is calculated as
\begin{align}
  s^{q}
  =\exp\left(i\int_{\text{2-cycle}}qB\right)
  =\exp\left(i\int_{\text{2-cycle}}dC\right)
  =1,
\end{align}
since $C$ is normalized as $\int_{\text{2-cycle}}dC\in \Zb$. So $s$ is a $q$-th root of $1$ and it shows that this $(B,C)$ system is a formulation of the 2-form $\Zb_q$ gauge field.

This 2-form $\Zb_q$ gauge field $(B,C)$ couples to the $\SU (N)$ gauge field as follows.  First, extend the $\SU(N)$ gauge field $A$ to a $\U (N)$ gauge field $\Acal$ whose field strength is $\Fcal:=d\Acal -i \Acal \wedge \Acal$.
We declare the $\lambda$ gauge transformation \eqref{2formgauge} of $\Acal$ as
\begin{align}
  \Acal \to \Acal + \lambda \id,
\end{align}
where $\id$ is the $N\times N$ identity matrix.  Then the trace part of the field strength transforms as
\begin{align}
  \tr(\Fcal)\to \tr(\Fcal)+Nd\lambda.
\end{align}
In order to gauge away the $\U (1)$ part of $\Acal$ we impose the constraint
\begin{align}
  \tr(\Fcal)-N B=0.
  \label{constrantFcal}
\end{align}
Notice that this constraint is $\lambda$ gauge invariant.  

Let us consider the instanton number $\nu$
defined in eq.~\eqref{ABJanomaly}. Actually $\nu$ is not an integer any more in the presence of the background $(B,C)$. Since $F$ is the traceless part of $\Fcal$, $\nu$ is calculated as
\begin{align}
  \nu
  &=\frac{1}{8\pi^2}\int \tr(F\wedge F)\nonumber\\
  &=\frac{1}{8\pi^2}\int \tr \left[\left(\Fcal-\frac{1}{N}\tr(\Fcal) \id\right)\wedge \left(\Fcal-\frac{1}{N}\tr(\Fcal)\id\right) \right]\nonumber\\  
  &=
  \frac{1}{8\pi^2}\int \tr \left[\Fcal\wedge \Fcal \right]
  -\frac{1}{8\pi^2 N}\int \tr(\Fcal)\wedge \tr(\Fcal).
  \label{calcnu1}
\end{align}
The first term in eq.~\eqref{calcnu1} is written as $\int c_2(\Fcal)+\frac12 \int c_1(\Fcal)\wedge c_1(\Fcal)$ and turn out to be an integer if the space-time is a spin manifold.  By using the constraint \eqref{constrantFcal}, we find
\begin{align}
  \nu=-\frac{N}{8\pi^2}\int B \wedge B\mod 1.
\end{align}
Taking eq.~\eqref{ABJanomaly} into account, the fermion path-integral measure changes by the transformation \eqref{chiralsymmetry} as
\begin{align}
  \int D\psi D\psib \to \int D\psi D\psib \, e^{2\pi i n \nu'},\quad
  \nu':=-\frac{N}{8\pi^2}\int B \wedge B.
  \label{mixedanomaly}
\end{align}
Actually $\frac{1}{2\pi}\int_{\text{2-cycle}}B$ takes value in $\frac{1}{q}\Zb$.  
Since our 4-dimensional space-time is a spin manifold and the intersection form is even, $\nu'$ takes value in 
\begin{align}
  \frac{N}{q^2} \Zb + \Zb
  =\frac{N_0'}{q'} \Zb + \Zb
  =\frac{1}{q'} \Zb,
\end{align}
where two relatively prime positive integers $N_0'$ and $q'$ are defined by the relation $N_0/q=N_0'/q'$.  Actually a configuration with $\nu=p/q' \ (p\in \Zb)$ is realized by a 4-torus with twisted boundary condition \cite{tHooft:1979rtg,tHooft:1981nnx,Witten:2000nv}.  As a result, the fermion path-integral measure changes by the transformation \eqref{chiralsymmetry} unless $n$ is multiple of $q'$.  In other words, chiral symmetry $\Zb_{\ell}$ is broken to $\Zb_{\ell/q'}$ in the presence of $(B,C)$ background.  This is the mixed 't~Hooft anomaly we are looking for.

This 't~Hooft anomaly is RG invariant by the same argument as \cite{tHooft:1979rat} and constrain the phase if $q'>1$ as follows.  
Assume the theory is in the confining phase, i.e.\ gapped and the center symmetry is not broken.  
Then the chiral symmetry must be broken at least to $\Zb_{\ell/q'}$ since otherwise the isolated vacuum preserve the symmetry larger than $\Zb_{\ell/q'}$ but the low energy effective theory is empty and cannot reproduce the 't Hooft anomaly.

In this argument we make the non-trivial assumption that a gauge theory with a simply connected gauge group does not have any topological order which reproduce the 't Hooft anomaly, if the theory is in the confining phase and the global symmetries are not spontaneously broken. Actually, it has been known that a class of 't Hooft anomalies can be produced by certain topological field theories \cite{Wang:2017loc,Tachikawa:2017gyf}.
It is not easy to fully justify this assumption and we postpone it to future work.

Notice that $c=0$ or $c=N/2$ as discussed in the last section. When $c=0$, $q=N$ and $q'=N$ as well.  On the other hand, when $c=N/2$, $q=N/2$ and $N_0=2$.  In this case $q'=N/4$ if $N$ is a multiple of $4$, and $q'=N/2$ if $N$ is an even number but not a multiple of $4$.  Therefore we have $q'>1$ unless $N=4$ and $c=2$.

\subsection{Domain-wall}
When a global discrete symmetry is spontaneously broken, the theory includes dynamical domain-walls connecting distinct vacua.  
Here let us briefly discuss these domain-walls, when the $\Zb_{\ell}$ chiral symmetry is spontaneously broken to $\Zb_{\ell/q'}$ in our theory.  
Such a domain-wall supports a conformal field theory or a topological field theory which reproduce the 't Hooft anomaly for the center symmetry $\Zb_q$ as discussed in \cite{Gaiotto:2014kfa}.

Let us see this 't Hooft anomaly.  Suppose the low energy theory is described by a periodic scalar field $\phi\sim \phi+2\pi$ which is transformed by the  $\Zb_{\ell}$ chiral symmetry transformation \eqref{chiralsymmetry} as
\begin{align}
  \phi \to \phi+ \frac{2\pi n}{q'}.
\end{align}
The $q'$ distinct vacua are parametrized as $\phi=\frac{2\pi m }{q'},\ (m=0,1,2,\cdots,q'-1)$.  Then in order to reproduce the mixed 't Hooft anomaly \eqref{mixedanomaly}, the effective action $S_{\mathrm{eff}}$ should include the coupling
to the background $(B,C)$ field
\begin{align}
  S_{\mathrm{eff}} \ni \frac{iN q'}{8\pi^2}\int \phi B\wedge B.
  \label{Seff}
\end{align}

Now let us consider the domain-wall connecting the vacua $\phi=\frac{2\pi m }{q'}$ and $\phi=\frac{2\pi (m+n)}{q'}$, whose world-volume is denoted by $M_3$.  The term \eqref{Seff} includes the world-volume action  
\begin{align}
  S_{\mathrm{eff}} \ni -\frac{iN n}{4\pi q^2}\int_{M_3}  C\wedge dC=-\frac{iN_0' n}{4\pi q'}\int_{M_3} C\wedge dC,
  \label{worldvolumeanomaly}
\end{align}
where we use the constraint \eqref{constraint}.  Therefore the partition function of the 3-dimensional worldvolume theory alone has the phase ambiguity due to the Chern-Simons term \eqref{worldvolumeanomaly} at fractional level, although the total theory does not have any phase ambiguity (see eq.~\eqref{Seff}).  This is nothing but the anomaly inflow mechanism \cite{Callan:1984sa}.  We can conclude that the domain-wall should support a conformal field theory or a topological field theory which reproduces this 't Hooft anomaly for 1-form $\Zb_{q'}\subset \Zb_{q}$ symmetry.  For example $N_0'n$ copies of the level $q'$ $U(1)$ Chern-Simons theory reproduce this 't Hooft anomaly \cite{Gaiotto:2014kfa}.

\section{Example of chiral symmetry breaking without bilinear condensate}
\label{sec:example}

\subsection{Example}
Here we consider the $\SU (6)$ gauge theory with a Weyl fermion in \yasym for an example. This representation \yasym is pseudo-real and thus there is no perturbative gauge anomaly.  The Dynkin index is $\ell=6$, and $6$-ality is $c=3$.  It would be useful to consult refs.~\cite{Slansky:1981yr,Yamatsu:2015npn} to find data of various representations.

Let us apply the result of section \ref{sec:anomaly} to this example.  We find $q=\gcd(N,c)=\gcd(6,3)=3$ and $N_0=N/q=2$ and therefore $q'=3$.  
As a result if the theory is in the confining phase, the chiral symmetry $\Zb_6$ is broken to $\Zb_2$. So there are 3 distinct vacua related by the broken elements in $\Zb_6/\Zb_2\cong \Zb_3$.

\subsection{Confinement}
The 't Hooft anomaly argument in this paper is powerful only when the theory is in the confining phase.  
Here we argue that our example of $\SU (6)$ with $R=$\yasym is quite likely to be in the confining phase.

Let us compare our theory with $\mathcal{N}=1$ $\SU (6)$ pure super Yang-Mills theory, which is known to be a confining theory.  
This pure super Yang-Mills theory is in the class of theories we considered in section \ref{sec:anomaly}.  
Actually this pure super Yang-Mills theory is the theory with $R=$adjoint.

We observe that $R=$\yasym is a ``smaller'' representation than $R=$adjoint in the dimensions, the Dynkin index, and the quadratic Casimir.  
See table \ref{tab:reps}.
This observation leads to the fact that the beta function coefficients of the theory with $R=$\yasym is larger than those of the theory with $R=$adjoint at least up to 3-loop in $\overline{\mathrm{MS}}$ scheme (see for example \cite{Herzog:2017ohr}).  
In other words the theory with $R=$\yasym becomes strongly coupled in the low energy more rapidly than the theory with $R=$adjoint.  
This observation combined with the fact that the theory with $R=$adjoint is in the confining phase strongly suggests that the theory with $R=$\yasym is also in the confining phase.

Combined with the result of section \ref{sec:anomaly}, it is quite likely that in the $\SU (6)$ gauge theory with \yasym, the chiral symmetry $\Zb_6$ is broken to $\Zb_2$ spontaneously.

\begin{table}
  \centering
  \begin{tabular}{|c|c|c|c|}\hline
      & dim & $\ell$ & $C_2$ \\\hline
    \yasym & 20 & 6 & 21/4 \\\hline
    adjoint & 35 & 12 & 6 \\\hline
  \end{tabular}
  \caption{The table of the number of dimensions dim, the Dynkin index $\ell$ and the quadratic Casimir $C_2$ of the representations \yasym and adjoint.  We observe that these quantities of the adjoint representation are larger than those of the \yasym.}
  \label{tab:reps}
\end{table}

\subsection{No fermion bilinear}
\label{sec:Nofermionbilinear}
So far, we have seen that the chiral symmetry $\Zb_6$ is spontaneously broken to $\Zb_2$. 
One may think that this spontaneous symmetry breaking is caused by the condensation of the fermion bilinear i.e.\ $\langle \psi\psi \rangle\ne 0$, but it cannot be true.  
Actually a Lorentz invariant and gauge invariant bilinear form in our theory identically satisfies
\begin{align}
  \psi\psi:=\epsilon^{\alpha\beta}\psi_{\alpha}^{I}\psi_{\beta}^{J}B_{IJ}=0,
\end{align}
since both the gauge invariant bilinear form $B_{IJ}$ and the Lorentz invariant bilinear form $\epsilon^{\alpha\beta}$ are anti-symmetric.
Therefore this bilinear form can never condense.

Let us show the explicit form of the gauge invariant $B_{IJ}$ in order to confirm that it is anti-symmetric.  
Since $I,J$ are the labels of the rank 3 anti-symmetric representation, they are expressed by the three anti-symmetrsized fundamental indices as
\begin{align}
  I=[i_1i_2i_3],\quad J=[j_1j_2j_3],\qquad (i_a,j_a=1,\dots,6).
\end{align}
By this notation, the gauge invariant bilinear form $B_{IJ}$ is expressed as
\begin{align}
  B_{IJ}=B_{[i_1i_2i_3][j_1j_2j_3]}:=\epsilon_{i_1i_2i_3j_1j_2j_3},\label{BIJ}
\end{align}
where $\epsilon_{i_1\dots j_3}$ is the totally anti-symmetric tensor, which is $\SU (6)$ gauge invariant.  From the expression \eqref{BIJ} we can confirm that $B_{IJ}$ is actually anti-symmetric.

We conclude that in the $\SU (6)$ gauge theory with a Weyl fermion in \yasym, the chiral symmetry $\Zb_6$ is spontaneously broken to $\Zb_2$,  although the fermion bilinear cannot condense.

\section{Summary and discussion}
\label{sec:summary}
In this paper we consider $\SU (N)$ gauge theory with a Weyl fermion in an irreducible self-conjugate representation $R$.  
We obtain constraints on the chiral symmetry breaking in the confining phase in this theory.  
These constraints are derived from 't~Hooft anomaly matching condition of the chiral symmetry and the center symmetry.
In particular in $\SU (6)$ gauge theory with a Weyl fermion in \yasym, the chiral symmetry $\Zb_6$ is spontaneously broken to $\Zb_2$ although the fermion bilinear is identically zero.

One natural question is what causes this spontaneous symmetry breaking if it is not the fermion bilinear.  One candidate is the four-fermi operator $\psi\psi\psi\psi$ which has the correct quantum number to break $\Zb_6$ to $\Zb_2$.  Actually there are two possible gauge invariant and Lorentz invariant 
$\psi\psi\psi\psi$ operators.  As we have seen, a Lorentz invariant fermion bilinear $\epsilon^{\alpha\beta}\psi_{\alpha}^{I}\psi_{\beta}^{J}$ must be symmetric in $I,J$.  The symmetric tensor product of \yasym is decomposed as
\begin{align}
  \ydiagram{1,1,1}\otimes_{\text{sym}}\ydiagram{1,1,1}=\ydiagram{2,1,1,1,1}\oplus \ydiagram{2,2,2}.
\end{align}
A gauge invariant $\psi\psi\psi\psi$ operator is given by the gauge invariant bilinear of each irreducible representation of the right-hand side.  It is an interesting future problem to find which four-fermi operator condenses.

Other candidates which causes this spontaneous symmetry breaking are fermion bilinear including derivatives.  
For example one may guess that $\langle \psi D_{\mu}D^{\mu} \psi \rangle \ne 0$ causes this spontaneous symmetry breaking.   
However one can show $\langle \psi D_{\mu}D^{\mu} \psi \rangle=0$ as follows.  
Since $\langle \psi D^{\mu} \psi \rangle=0$ due to the Lorentz symmetry, the derivative of this equation reads
\begin{align*}
  0=\del_{\mu}\langle \psi D^{\mu} \psi \rangle=\langle D_{\mu}\psi D^{\mu} \psi + \psi D_{\mu}D^{\mu} \psi \rangle=\langle \psi D_{\mu}D^{\mu} \psi \rangle,
\end{align*}
where we use $D_{\mu}\psi D^{\mu} \psi=0$ shown by the same argument as in subsection \ref{sec:Nofermionbilinear}.  
Therefore a fermion bilinear which may condense must include four or more derivatives.

Our theory, $\SU (6)$ gauge theory with a Weyl fermion in \yasym, does not have any known gauge anomaly.  
However it is not guaranteed to be a consistent theory, since there is possibility that it is suffered from some unknown gauge anomaly.  
In fact, our theory cannot have fermion mass term and this fact may suggest that our theory has some gauge anomaly.  
For example recently a new global anomaly is found by \cite{Wang:2018qoy}.  
One way to guarantee that a theory is anomaly free is to construct a gauge invariant and non-perturbatively regularized lattice gauge theory.  
This is also a challenging problem.

Interestingly there is a string theory realization of the $\mathcal{N}=2$ $\SU (6)$  supersymmetric gauge theory with a half hypermultiplet in \yasym 
\cite{Tachikawa:2011yr}.  By adding mass term to all the scalars and adjoint fermions in this theory, we obtain the $\SU (6)$ gauge theory with a Weyl fermion in \yasym.  This fact may suggest that our theory is free from any gauge anomaly and a consistent gauge theory.

One may be curious whether there are other examples in which the chiral symmetry is spontaneously broken without fermion bilinear condensate.  
Actually, $\SU(4k+2),\ k=1,2,3,\dots$ gauge theory with a Weyl fermion in rank $(2k+1)$ anti-symmetric representation has similar 't Hooft anomaly and a vanishing fermion bilinear.  
However their beta-functions are quite different.  For example $k=2,\ \SU(10)$ case, the rank 5 anti-symmetric representation is quite large, and the confinement is not quite likely to occur.  
Actually the Caswell-Banks-Zaks fixed point \cite{Caswell:1974gg,Banks:1981nn} is found in the weakly coupled regime by looking at the 2-loop beta function in this theory.  
For $k>2$ the theory is not asymptotically free any more.

There are a lot of interesting future problems. 
For example, we can consider more general gauge theories including many flavors as well as different irreducible representations.  
It will also be an interesting problem to investigate finite temperature phase transitions.  

\subsection*{Acknowledgement}
I would like to thank Prarit Agarwal, Hidenori Fukaya, Dongmin Gang, Yuta Hamada, Naoki Kawai, Seok Kim, Yoshiyuki Matsuki,  Tetsuya Onogi, Shigeki Sugimoto, Yuji Tachikawa, Seiji Terashima, Mithat \"{U}nsal, and Shuichi Yokoyama for discussions and comments.
I would also thank the Yukawa Institute for Theoretical Physics at Kyoto University. 
Discussions during the YITP workshop YITP-W-17-08 on ``Strings and Fields 2017'' and YITP-W-18-04 on ``New Frontiers in String Theory 2018'' were useful to complete this work. 
This work was supported in part by JSPS KAKENHI Grant Number 15K05054.

\bibliographystyle{utphys}
\bibliography{refs}
\end{document}